# Comparative Analysis of RNA Families Reveals Distinct Repertoires for Each Domain of Life

Marc P. Hoeppner[a], Paul P. Gardner[b] & Anthony M. Poole[b]*

[a]Science for Life Laboratory, Department of Medical Biochemistry and Microbiology, Uppsala University, SE-751 23 Uppsala, Sweden

[b]School of Biological Sciences, University of Canterbury, Private Bag 4800, Christchurch 8140, New Zealand

*To whom correspondence should be addressed.

Email addresses:

anthony.poole@canterbury.ac.nz

paul.gardner@canterbury.ac.nz

mphoeppner@gmail.com

**Running head: RNA Repertoires of the Three Domains of Life**

## Abstract

The RNA world hypothesis, that RNA genomes and catalysts preceded DNA genomes and genetically-encoded protein catalysts, has been central to models for the early evolution of life on Earth. A key part of such models is continuity between the earliest stages in the evolution of life and the RNA repertoires of extant lineages. Some assessments seem consistent with a diverse RNA world, yet direct continuity between modern RNAs and an RNA world has not been demonstrated for the majority of RNA families, and, anecdotally, many RNA functions appear restricted in their distribution. Despite much discussion of the possible antiquity of RNA families, no systematic analyses of RNA family distribution have been performed. To chart the broad evolutionary history of known RNA families, we performed comparative genomic analysis of over 3 million RNA annotations spanning 1446 families from the Rfam 10 database. We report that 99% of known RNA families are restricted to a single domain of life, revealing discrete repertoires for each domain. For the 1% of RNA families/clans present in more than one domain, over half show evidence of horizontal gene transfer, and the rest show a vertical trace, indicating the presence of a complex protein synthesis machinery in the Last Universal Common Ancestor (LUCA) and consistent with the evolutionary history of the most ancient protein-coding genes. However, with limited interdomain transfer and few RNA families exhibiting demonstrable antiquity as predicted under RNA world continuity, our results indicate that the majority of modern cellular RNA repertoires have primarily evolved in a domain-specific manner.

## Author Summary

In cells, DNA carries recipes for making proteins, and proteins perform chemical reactions, including replication of DNA. This interdependency raises questions for early evolution, since one molecule seemingly cannot exist without the other. A resolution to this problem is the RNA world, where RNA is postulated to have been both genetic material and primary catalyst. While artificially selected catalytic RNAs strengthen the chemical plausibility of an RNA world, a biological prediction is that some RNAs should date back to this period. In this study, we ask to what degree RNAs in extant organisms trace back to the common ancestor of cellular life. Using the Rfam RNA families database, we systematically screened genomes spanning the three domains of life (Archaea, Bacteria, Eukarya) for RNA genes, and examined how far back in evolution known RNA families can be traced. We find that 99% of RNA families are restricted to a single domain. Limited conservation within domains implies ongoing emergence of RNA functions during evolution. Of the remaining 1%, half show evidence of horizontal transfer (movement of genes between organisms), and half show an evolutionary history consistent with an RNA world. The oldest RNAs are primarily associated with protein synthesis and export.

## Introduction

Following demonstration that RNA can act as genetic material [1-3] and biological catalyst [4,5], the study of the origin and early evolution of life on Earth has been heavily focused on the potential for an RNA world. The RNA world hypothesis is that RNA was both genetic material and main biological catalyst, prior to the advent of DNA and templated protein synthesis [6-8]. The chemical plausibility of an RNA world has been intensively investigated through the application of in vitro methodologies that enable selection and subsequent characterization of novel RNA functionalities [9,10]. Equally, the discovery of naturally-occurring functional RNAs in biological systems has expanded our understanding of the ways in which extant organisms utilize this macromolecule in a wide range of contexts, including catalysis, regulation, and as sequence-based guides [11-15].

A central tenet of RNA world theory as an account of the early evolution of life on Earth is the Principle of Continuity [6], namely, that modern systems are the product of gradual evolution from earlier states. Consequently, it is possible that some RNA families could be direct descendents of molecules that first evolved in the RNA world [16,17]. The broad functionality of RNA both in terms of catalysis and biological function hints at a possibly complex RNA world [12,17,18], but assessing the antiquity of individual RNA families has been hampered by limited comparative data, and difficulties in annotating RNAs in genomes [19]. At the same time, it seems likely that many RNA families significantly postdate the RNA world, having evolved de novo much later in the evolution of life [13,20]. Indeed, for protein-coding genes, both very deep evolutionary histories [21-23] and more recent origins [24,25] have been established.

Assigning relic status to individual RNAs is not without significant complication. First, placing RNAs with non-universal distributions into the common ancestor of archaea, bacteria and eukaryotes requires lineage or domain-specific losses to be invoked [26]. While loss is plausible, it is difficult to verify at the level of cellular domains, since recent origin versus lineage-specific loss following a more ancient origin cannot be readily distinguished, and other data must be considered [27]. Another process that may obfuscate the history of early RNA-based life is the propensity for genes to undergo horizontal transmission, from a donor to a recipient. For protein-coding genes, there is now overwhelming evidence that horizontal gene transfer is a significant evolutionary force, particularly for microbes [28,29]. Consequently, gene-based phylogenies do not always provide an accurate means of gauging the evolutionary history of species, and, extrapolating across the tree of life and several billion years of evolutionary history, it is plausible that no gene will have remained untouched by horizontal gene transfer [30]. Consequently, historical signal consistent with RNA world continuity may have been erased through subsequent gene transfer events. Conversely, effective spread by horizontal transmission could lead to RNAs appearing artificially ancient. Finally, many RNAs may be more recent evolutionary innovations, and may not be RNA world relics [13].

These concerns notwithstanding, it remains commonplace for novel RNAs or RNA families to be discussed in regard to their potential relevance to the RNA world. Indeed, there are countless qualitative surveys derived from review of the experimental literature (see for example [11,12,14,17,18,31]), which often extrapolate deep evolutionary origins from limited comparative data. Problematically, this approach has led to RNA world model being populated with RNAs whose distributions are patchy, and antiquity has often been inferred on speculative grounds, following detailed experimental characterisation of RNAs from a handful of model organisms. Against this backdrop, it is perhaps of little surprise that more vociferous critics have dubbed this endeavour the 'RNA dreamtime' [32].

While detailed studies have been performed for single RNA families (Table S1), no published data present a systematic analysis covering all RNA families, despite this now being routine for protein-coding genes. For RNA genes, an equivalent analysis is long overdue but has not been possible because, until recently, comparative data were not of sufficiently high quality.

We therefore sought to systematically address whether the phylogenetic distribution of extant RNAs fits with direct descent from an RNA world, as predicted under the Continuity hypothesis, or whether the distribution of extant RNAs better reflects more recent (post-LUCA) origins. In addition, we sought to examine whether horizontal transfer between cellular domains (and viruses) is detectable for RNA families. We report an analysis of over 3 million RNAs spanning 1446 families in the Rfam database [33], revealing that the overwhelming majority of families (99%) are restricted to a single domain of life. By contrast, fewer than 1% show evidence of either a deeper evolutionary origin, or of interdomain transfers. We conclude that, while, on these proportions, the RNA world 'palimpsest' is only a fraction of the RNA repertoires of modern genomes, the most ancient RNA families nevertheless belie evidence of an advanced protein synthesis apparatus. Strikingly, we report that interdomain horizontal gene transfers are also minimal for RNA genes, in marked contrast to the significant levels detected for protein-coding genes. Our analyses thus serve to move the current state-of-the-art from erudite literature review to systematic analysis of the distribution and antiquity of large numbers of RNA families.

## Results/Discussion

**99% of RNA families are restricted to a single domain of life.**

We first asked whether a systematic analysis of RNA families expands our knowledge of ancient RNAs beyond those identified by traditional experimental work. To examine the degree to which extant RNAs can be traced to earlier evolutionary periods, we performed comparative analyses of annotated RNAs based on data from all three domains of life as well as viruses. To this end, we used the Rfam (RNA families) database [33], which groups RNAs into families, and families into clans, based on manually-curated alignments, consensus secondary structures, covariance models [34] and functional annotations. RNAs within families and clans can therefore be claimed to share a common ancestry [33]. All analyses presented here are based on Rfam 10.0, which consists of over 3 million annotations grouped into 1446 families and 99 clans [33].

To generate a high-quality dataset, we first established the distribution of all individual RNA sequence entries in Rfam by reference to the NCBI taxonomy database, and manually vetted and removed probable false positive annotations. From the resulting dataset, we generated an initial survey of families and clans across bacterial, archaeal, eukaryotic and viral genomes (**Fig. 1**). Two patterns are immediately clear. First, each domain carries a large number of entries absent from the other domains, with limited overlap observed between domains, or with viruses. Second, only seven Rfam families are present across all three domains. That we observe distinct domain-level RNA repertoires appears consistent with the view that the three domains of life are genetically distinct [35]. However, families present in more than one domain (or shared with viruses) may be the result of either vertical evolution from a common ancestor or horizontal transfer of genes between domains [29,35].

**Interdomain RNA families show a mix of vertical and horizontal inheritance.** We next sought to establish whether the distribution the 12 interdomain Rfam families/clans (**Fig. 1**) could be attributed either to vertical inheritance or horizontal gene transfer. Previous studies and data on distribution allow a predominantly vertical pattern of inheritance to be attributed to only five families (SSU and 5S rRNAs, tRNA, RNase P RNA, SRP RNA) with four showing evidence of HGT (group I & II introns, organellar LSU rRNA, IsrR RNA) (**Table S1**). Ribosomal RNAs are not fully represented in Rfam, being amply covered by other databases (e.g. [36,37]), but their deep evolutionary history has been readily traced (Table S1). Combined, these data confirm a minimal reconstruction of the RNA repertoire of LUCA consistent with that observed for protein-coding genes [21], with the demonstrably oldest RNAs and the majority of such proteins being involved in translation and protein export (**Fig. 2**). Consequently, while the number of RNA families traceable to LUCA is an order of magnitude lower than for proteins, the spread of functionalities is nevertheless very similar in extent.

A vertical trace is suspected but not demonstrated for the universally distributed TPP riboswitch (**Table S1, Fig. 3**), which modulates gene expression in response to thiamine pyrophosphate (TPP). The analysis of patterns of inheritance for RNAs is complicated by their short lengths and generally low levels of sequence conservation. As riboswitches regulate cognate mRNA in cis, vertical transmission may be tested by generating phylogenies from the protein products, on the assumption that the riboswitch and ORF have coevolved. We therefore generated a phylogeny for THIC, the only TPP-regulated gene product present in all three domains. The phylogeny shows eukaryote sequences grouping with proteobacteria (**Fig. S1**), consistent with horizontal transmission of TPP-riboswitch regulated ThiC to the eukaryote lineage from a bacterial donor. Several independent observations are consistent with horizontal transmission: *Arabidopsis* THIC is nuclear-encoded, but targets to the chloroplast [38], plant ThiC can complement an *E. coli* ThiC mutant [39], and eukaryotic TPP riboswitches show limited distribution [40] (Rfam 10.0). Moreover, THI1, which also carries a TPP riboswitch in its mRNA leader, is also targeted to chloroplasts and mitochondria [41]. While an early origin for TPP riboswitches [11] remains plausible, this is difficult to reconcile with our THIC phylogeny, since bacterial and archaeal sequences are not monophyletic under any rooting (**Fig. S1**).

Also noteworthy is the CRISPR/Cas system, which combats viral and plasmid infection in both bacteria and archaea. Horizontal transmission has been suggested for this system, but interdomain transfer is thought to be limited [42]. Examination of CRISPR crRNA family distribution reveals that 54 of 65 Rfam crRNA families are restricted to a single domain (**Table S2**). The remaining 11 families fall into two clans (CRISPR-1, CRISPR-2), which include crRNAs in both bacterial and archaeal genomes. However, only one Rfam family from each of these two clans contains annotations deriving from both domains. While short sequence length of crRNAs precludes phylogenetic analyses, the distribution we report (**Table S2**) is compatible with sporadic interdomain transfer, consistent with a phylogenomic analysis of Cas genes/clusters which reported low levels of horizontal transmission [43].

The low number of observed interdomain RNA families suggests that, in contrast to protein-coding gene repertoires, RNA repertoires are surprisingly refractory to interdomain transfers. While we do see evidence of organellar contributions, these are few in number, in marked contrast to the high numbers observed for protein-coding genes [44,45].

**Only a minority of domain-specific RNA families are broadly-distributed.** We next sought to establish the distribution of RNA families within each domain, since our initial analysis (**Fig. 1**) does not consider within-domain taxonomic distribution of Rfam families. A broad distribution may indicate an early origin of a given family, but information on distribution alone cannot distinguish between horizontal and vertical modes of transmission. As short length and limited sequence conservation preclude robust phylogenies for the vast majority of RNA families, distribution cannot be used to directly infer the RNA repertoire of the last common ancestor (LCA) of each domain. Nevertheless, such information may indicate whether the RNA repertoires of the three domains are functionally distinct. We therefore collated families present in at least 50% of major within-domain taxonomic divisions (**Fig. 3, Data S2**). Surprisingly, the number of broadly distributed families/clans within each domain is small (Archaea 13/69=18.8%, Bacteria 15/223=6.7%, Eukaryotes 20/826=2.4%), though among eukaryotes there are a high number of clans, which may encompass multiple RNA families with a shared evolutionary history. Two patterns emerge from this analysis (**Fig. 3**). First, eukaryote and archaeal repertoires are dominated by snoRNAs. Second, the most broadly distributed bacterial RNAs are regulatory.

Closer investigation of the snoRNA repertoires across archaea and eukaryotes reveals that C/D family RNAs are broadly distributed; H/ACA family RNAs, while widespread among eukaryotes, are only known from Euryarchaeota [46,47], and Archaeal H/ACA RNAs are not currently included in Rfam [33]. Strikingly, of the >500 snoRNA families included in this study, none are shared across archaea and eukaryotes. While a deep origin of snoRNPs is supported by surveys of protein and RNA components [48], this is not reflected by existence of conserved RNA families, for which only scant evidence exists [49,50].

In eukaryotes, a strong domain-specific evolutionary trace is attributable to snRNAs (**Fig. 3, Table S3**), consistent with other studies indicating both the major and minor spliceosome were features of the Last Eukaryotic Common Ancestor (LECA) [51-53].

A different picture emerges for miRNAs however. The broad distribution of miRNAs is consistent with the suggestion that RNAi pathways trace to the LECA [54], with 26/452 miRNA families present in more than one eukaryotic supergroup (**Data S3**). However, closer inspection reveals most are singleton false positives or artefactual family groupings. Our dataset therefore does not allow the placement of any individual miRNA families in LECA.

A broad qualitative difference between bacteria compared to archaea and eukaryotes is the preponderance of conserved regulatory elements, primarily riboswitches (**Fig. 3**). However, this observation is based on only that small fraction of Rfam families present in ≥50% of taxonomic divisions. To further assess whether there are qualitative differences between the functional RNA repertoires across the three domains and viruses, we took advantage of the organization of Rfam into different functionalities. As is evident from **Fig. 4**, common functionalities across all three domains are sparse. Riboswitches and ribozymes indicate the ubiquity of small metabolite-based regulation and catalytic function, but of the numerous families included in this analysis, only RNase P RNA is directly traceable to the LUCA (**Figs. 2 & 3**). Functionalities shared between archaea and eukaryotes to the exclusion of bacteria are restricted to snoRNA-dependent RNA modification, and CRISPRs are the only prokaryote-specific functionality. Interestingly, a number of RNA functionalities present in bacteria lack archaeal or eukaryotic representatives (cis-regulatory leaders, thermoregulators, sRNAs), and Rfam contains no archaeal-specific functionalities (**Fig. 4, Data S4**), possibly attributable to the smaller number of experimental screens for novel RNAs across members of this domain.

**Biases in taxonomic sampling.** In comparing the RNA repertoires of the three domains, a key question is whether the underlying Rfam data cover a reasonable spread of species within each domain, or whether data from a few species or phyla dominate. This is important in that the low number of broadly distributed families/clans we observe within each domain could be the result of an underlying sampling bias. A priori we may expect a significant bias, given current genomic coverage of microbial biodiversity. For instance, a recent survey of snoRNAs indicates there is broad, though nevertheless patchy coverage across major eukaryotic and archaeal groups [48]. We therefore examined the underlying taxonomic distribution of all domain-specific Rfams. For all three domains, entries are heavily skewed, with a majority of Rfam annotations deriving from a narrow phylogenetic diversity (**Fig. S2**).

For protein-coding genes, discovery of novel proteins has been significantly enhanced by sequencing of genomes chosen for maximal phylogenetic diversity [55]. While de novo computational discovery of novel ncRNAs is non-trivial by comparison, we were nevertheless interested in establishing whether the additional phylogenetic coverage provided by the Genomic Encyclopedia of Bacteria and Archaea (GEBA) [55] impacted the number of broadly distributed Rfam families. Under the assumption of vertical inheritance, we therefore treated RNAs as characters on the GEBA phylogeny. Our analysis yielded four additional bacterial candidates (marked with asterisks in **Fig. 3**), though again we caution that broad distribution may be generated through HGT, so these candidates cannot be placed in the bacterial ancestor. Nevertheless, this modest improvement suggests GEBA [55], and targeted experimental screens informed by phylogeny [48] will provide a valuable framework, both for improving knowledge of RNA family distribution and in focusing experimental screens for novel RNA families.

How should we interpret these data? The limited distribution of domain-specific RNAs is likely to be biased by sampling, a problem that affects all genomic data, and is even more acute for detailed experimental data. On available data, we find that only a minority of domain-specific RNAs exhibit a broad distribution. A broad distribution could result from vertical inheritance, but it could also be the result of horizontal gene transfer. Taxonomic biases might underestimate the number of RNAs vertically traceable to the ancestor of a domain, whereas horizontal gene transfer might be expected to expand the distribution of some RNAs. Assuming that current sampling has gaps, but is not completely uninformative [48], available data suggest that a high proportion of RNAs are likely to be evolutionarily young, and will not trace to the LCA of the domain in which they reside.

**Concluding remarks.** We have examined the evolution and diversity of RNAs across the entire tree of life, an important complement to previous comparative studies on RNA metabolism [11,17] and RNA-associated protein families [56]. Large-scale analyses of the RNA repertoire are only now becoming possible through improved methodologies for RNA identification and greater integration between RNA discovery and online databases.

It is commonplace for novel RNAs or RNA families to be discussed in regard to their potential relevance to the RNA world, yet RNAs with limited distribution are difficult to reconcile with a very ancient evolutionary origin unless massive losses are invoked. Excepting the possibility of losses (which cannot be readily tested since the evidence for antiquity has been erased), our study shows that direct evidence for the RNA continuity hypothesis remains scant; there is undoubtedly an RNA 'palimpsest' [16], but it is not possible to expand this through systematic comparative analyses.

Conversely, we find clear evidence of distinct domain-level repertoires, but limited evidence of inter-domain transfers, consistent with a recent analysis indicating a detectable vertical trace amidst ongoing HGT [29]. The paucity of shared eukaryotic and archaeal RNA regulatory processes (**Fig. 4**) and the marginal bacterial contribution to the eukaryote RNA repertoire, support the view that eukaryotic mechanisms of RNA regulation are a domain-specific invention [15], and extend this view to the other two domains. While we see qualitative similarities between archaea and eukaryotes (**Figs. 3 & 4**), in agreement with studies indicating a phylogenetic affinity between these two domains [57], these are currently restricted to snoRNAs. The clear differences in RNA functional repertoires between eukaryotes, archaea and bacteria (**Fig. 4**) strengthen the case for recognizing the *biological* distinctness of the three domains [35], independent of uncertainty surrounding their specific phylogenetic relationships [58].

## Materials and Methods

**Rfam dataset.** Annotated noncoding RNA data used in this study was derived from data curated in Release 10.0 of the Rfam database [33] (http://rfam.sanger.ac.uk/). The distribution of Rfam families (**Data S1**) was established in two steps. First, for a given family, all annotations across the EMBL database [59] (http://www.ebi.ac.uk/embl/) were binned into domains using the taxonomic information attached to each sequence. We then inspected annotations from families whose distribution spanned more than one domain to identify possible false annotations. For all Rfam families with annotations spanning two or more domains (including viruses) we first confirmed the taxonomic affiliation of each sequence through reciprocal blasts against the GenBank database and removed any cases where sequences were clearly misannotated (e.g. bacterial sequencing vectors in eukaryote genome projects). Next, we inspected the quality of each annotation with reference to Rfam seed alignments. Any sequences with a bitscore within +10 bits of the individual bitscore cutoffs for curated seed alignments, and where sequence similarity was deemed insufficient to reliably establish homology, were discarded.

**Higher-level taxonomic assignments.** In assigning Rfam entries to specific taxonomic groups of bacteria and archaea (**Figure 3, Data S1**), we used the top-level classifications within each domain in the NCBI Taxonomy Database. At the time the analyses were performed, the proposed archaeal phylum Thaumarchaeota [60] was not recognised in the database, and available sequences were classified as Crenarchaeota. While members of the Thaumarchaeota are present in our data, none carry annotated snoRNAs, so not explicitly recognizing putative Thaumarchaeotes as a phylum does not impact the results summarized in figure 2. For Eukaryote RNA sequences, data was grouped according to the classification scheme proposed by Adl and colleagues [61].

**Phylogenetic analyses.** All sequences annotated as THIC in Genbank were retrieved (8 Feb 2011). The resulting list of 4508 sequences were examined for sequence similarity by generating a blast network using the blastall program from the BLAST package (version 2.2.18), with an E-value cutoff of 0.1. The network of blast results was visualized with CLANS [62], using default settings. The output was then clustered using

MCL [63], with granularity set at 4. Representative sequences spanning all domains were retrieved from all MCL clusters with >10 members. Sequences were aligned using MSA-Probs [64]. Partial sequences and extremely divergent sequences where annotation appeared questionable were removed. Conserved regions were selected for use in phylogenetic analysis via the G-blocks server [65] (http://molevol.cmima.csic.es/castresana/Gblocks_server.html), with the settings 'Allow smaller final blocks' and 'Allow gap positions within the final blocks' selected. ProtTest [66] was used to identify the best-fit model of protein evolution for our alignment. Phylogenetic analysis was performed using PhyML 3.0 [67] with parameters and model (WAG+I+G) as selected using ProtTest. Bootstrapping was performed on two Mac Pro machines with Intel Xeon Quad core processors, running 12 parallel threads. Parallelization yielded a total of 108 bootstrap replicates (a consequence of running 12 threads in parallel, resulting in bootstrap replicates that were a multiple of 12); all bootstrap values in figure S1 are therefore out of a total of 108 not 100. Additional trees were generated using RAxML [68] and BioNJ [69] to assess robustness of the topology. Tree figures were generated in Dendroscope [70].

## Abbreviations

LUCA — Last Universal Common Ancestor; Rfam — RNA families database; HGT — Horizontal Gene Transfer; SSU rRNA — Small Subunit ribosomal RNA; LSU rRNA — Large Subunit ribosomal RNA; SRP RNA — Signal Recognition Particle RNA; TPP riboswitch — Thiamine Pyrophosphate riboswitch; LCA — Last Common Ancestor;

snoRNA — small nucleolar RNA; GEBA — Genomic Encyclopedia of Bacteria and Archaea

## Competing interests

The author(s) declare that they have no competing interests.

## Authors' contributions

MPH and AMP conceived and designed research. MPH performed research, with AMP and PPG contributing additional analyses. AMP and MPH wrote the manuscript. All authors read and approved the final manuscript.

## Additional files

The following additional data are available with the online version of this paper.

**Supplementary file 1.** PDF with supporting text and references, supplementary figures S1-S3, and supplementary tables S1-S4.

**Additional Data file 1.** Distribution of Rfam families across domains and major phylogenetic groups (.xls file)

**Additional Data file 2.** Distribution of archaeal, eukaryote and bacterial Rfams (.xls file)

**Additional Data file 3.** Distribution of eukaryotic miRNAs in Rfam (.xls file)

**Additional Data file 4.** Numbers and taxonomic sources of annotations associated with RNA functional groups (.xls file)

## Acknowledgements

We thank D. Jeffares and J. Tylianakis for valuable comments on the manuscript.

## References

1. Fraenkel-Conrat H (1956) The role of the nucleic acid in the reconstitution of active Tobacco Mosaic Virus. Journal of the American Chemical Society 78: 882-883.
2. Gierer A, Schramm G (1956) Infectivity of ribonucleic acid from Tobacco Mosaic Virus. Nature 177: 702-703.
3. Diener TO (1971) Potato spindle tuber "virus". IV. A replicating, low molecular weight RNA. Virology 45: 411-428.
4. Kruger K, Grabowski PJ, Zaug AJ, Sands J, Gottschling DE, et al. (1982) Self-splicing RNA: autoexcision and autocyclization of the ribosomal RNA intervening sequence of Tetrahymena. Cell 31: 147-157.
5. Guerrier-Takada C, Gardiner K, Marsh T, Pace N, Altman S (1983) The RNA moiety of ribonuclease P is the catalytic subunit of the enzyme. Cell 35: 849-857.
6. Orgel LE (1968) Evolution of the genetic apparatus. J Mol Biol 38: 381-393.
7. Crick FH (1968) The origin of the genetic code. J Mol Biol 38: 367-379.
8. Gilbert W (1986) The RNA world. Nature 319: 618.
9. Joyce GF (2007) Forty years of in vitro evolution. Angewandte Chemie (International ed 46: 6420-6436.
10. Chen X, Li N, Ellington AD (2007) Ribozyme catalysis of metabolism in the RNA world. Chemistry & biodiversity 4: 633-655.
11. Breaker RR (2010) Riboswitches and the RNA World. Cold Spring Harb Perspect Biol: 10.1101/cshperspect.a003566.
12. Cech TR (2009) Crawling out of the RNA world. Cell 136: 599-602.
13. Eddy SR (2001) Non-coding RNA genes and the modern RNA world. Nat Rev Genet 2: 919-929.
14. Collins LJ, Kurland CG, Biggs P, Penny D (2009) The modern RNP world of eukaryotes. J Hered 100: 597-604.
15. Amaral PP, Dinger ME, Mercer TR, Mattick JS (2008) The eukaryotic genome as an RNA machine. Science 319: 1787-1789.
16. Benner SA, Ellington AD, Tauer A (1989) Modern metabolism as a palimpsest of the RNA world. Proc Natl Acad Sci USA 86: 7054-7058.
17. Jeffares DC, Poole AM, Penny D (1998) Relics from the RNA world. J Mol Evol 46: 18-36.
18. Yarus M (2002) Primordial Genetics: Phenotype of the Ribocyte. Annu Rev Genet 36: 125-151.
19. Freyhult EK, Bollback JP, Gardner PP (2007) Exploring genomic dark matter: a critical assessment of the performance of homology search methods on noncoding RNA. Genome Res 17: 117-125.
20. Mattick JS, Gagen MJ (2001) The evolution of controlled multitasked gene networks: the role of introns and other noncoding RNAs in the development of complex organisms. Mol Biol Evol 18: 1611-1630.

21. Harris JK, Kelley ST, Spiegelman GB, Pace NR (2003) The genetic core of the universal ancestor. Genome Res 13: 407-412.
22. Chothia C, Gough J, Vogel C, Teichmann SA (2003) Evolution of the protein repertoire. Science (New York, NY 300: 1701-1703.
23. Wang M, Yafremava LS, Caetano-Anolles D, Mittenthal JE, Caetano-Anolles G (2007) Reductive evolution of architectural repertoires in proteomes and the birth of the tripartite world. Genome research 17: 1572-1585.
24. Keese PK, Gibbs A (1992) Origins of genes: "big bang" or continuous creation? Proceedings of the National Academy of Sciences of the United States of America 89: 9489-9493.
25. Choi IG, Kim SH (2006) Evolution of protein structural classes and protein sequence families. Proceedings of the National Academy of Sciences of the United States of America 103: 14056-14061.
26. Penny D, Poole A (1999) The nature of the last universal common ancestor. Curr Opin Genet Dev 9: 672-677.
27. Penny D, Hoeppner MP, Poole AM, Jeffares DC (2009) An Overview of the Introns-First Theory. J Mol Evol.
28. Olendzenski L, Gogarten JP (2009) Evolution of genes and organisms: the tree/web of life in light of horizontal gene transfer. Annals of the New York Academy of Sciences 1178: 137-145.
29. Puigbo P, Wolf YI, Koonin EV (2009) Search for a 'Tree of Life' in the thicket of the phylogenetic forest. J Biol 8: 59.
30. Bapteste E, O'Malley MA, Beiko RG, Ereshefsky M, Gogarten JP, et al. (2009) Prokaryotic evolution and the tree of life are two different things. Biology Direct 4: 34.
31. Poole AM, Jeffares DC, Penny D (1998) The path from the RNA world. J Mol Evol 46: 1-17.
32. Kurland CG (2010) The RNA dreamtime: modern cells feature proteins that might have supported a prebiotic polypeptide world but nothing indicates that RNA world ever was. BioEssays 32: 866-871.
33. Gardner PP, Daub J, Tate J, Moore BL, Osuch IH, et al. (2011) Rfam: Wikipedia, clans and the "decimal" release. Nucleic Acids Res 39: D141-145.
34. Nawrocki EP, Kolbe DL, Eddy SR (2009) Infernal 1.0: inference of RNA alignments. Bioinformatics 25: 1335-1337.
35. Woese CR (2002) On the evolution of cells. Proc Natl Acad Sci USA 99: 8742-8747.
36. Cannone JJ, Subramanian S, Schnare MN, Collett JR, D'Souza LM, et al. (2002) The comparative RNA web (CRW) site: an online database of comparative sequence and structure information for ribosomal, intron, and other RNAs. BMC Bioinformatics 3: 2.
37. Pruesse E, Quast C, Knittel K, Fuchs BM, Ludwig W, et al. (2007) SILVA: a comprehensive online resource for quality checked and aligned ribosomal RNA sequence data compatible with ARB. Nucleic Acids Res 35: 7188-7196.
38. Raschke M, Burkle L, Muller N, Nunes-Nesi A, Fernie AR, et al. (2007) Vitamin B1 biosynthesis in plants requires the essential iron sulfur cluster protein, THIC. Proc Natl Acad Sci USA 104: 19637-19642.
39. Kong D, Zhu Y, Wu H, Cheng X, Liang H, et al. (2008) AtTHIC, a gene involved in thiamine biosynthesis in Arabidopsis thaliana. Cell Res 18: 566-576.
40. Sudarsan N, Barrick JE, Breaker RR (2003) Metabolite-binding RNA domains are present in the genes of eukaryotes. RNA 9: 644-647.
41. Chabregas SM, Luche DD, Van Sluys MA, Menck CF, Silva-Filho MC (2003) Differential usage of two in-frame translational start codons regulates subcellular localization of Arabidopsis thaliana THI1. J Cell Sci 116: 285-291.
42. Shah SA, Garrett RA (2011) CRISPR/Cas and Cmr modules, mobility and evolution of adaptive immune systems. Res Micro 162: 27-38.

43. Haft DH, Selengut J, Mongodin EF, Nelson KE (2005) A guild of 45 CRISPR-associated (Cas) protein families and multiple CRISPR/Cas subtypes exist in prokaryotic genomes. PLoS Comput Biol 1: e60.
44. Esser C, Ahmadinejad N, Wiegand C, Rotte C, Sebastiani F, et al. (2004) A genome phylogeny for mitochondria among alpha-proteobacteria and a predominantly eubacterial ancestry of yeast nuclear genes. Molecular Biology and Evolution 21: 1643-1660.
45. Martin W, Rujan T, Richly E, Hansen A, Cornelsen S, et al. (2002) Evolutionary analysis of Arabidopsis, cyanobacterial, and chloroplast genomes reveals plastid phylogeny and thousands of cyanobacterial genes in the nucleus. Proc Natl Acad Sci USA 99: 12246-12251.
46. Tang TH, Bachellerie JP, Rozhdestvensky T, Bortolin ML, Huber H, et al. (2002) Identification of 86 candidates for small non-messenger RNAs from the archaeon Archaeoglobus fulgidus. Proc Natl Acad Sci USA 99: 7536-7541.
47. Muller S, Leclerc F, Behm-Ansmant I, Fourmann JB, Charpentier B, et al. (2008) Combined in silico and experimental identification of the Pyrococcus abyssi H/ACA sRNAs and their target sites in ribosomal RNAs. Nucleic Acids Res 36: 2459-2475.
48. Gardner PP, Bateman A, Poole AM (2010) SnoPatrol: how many snoRNA genes are there? J Biol 9: 4.
49. Gaspin C, Cavaille J, Erauso G, Bachellerie JP (2000) Archaeal homologs of eukaryotic methylation guide small nucleolar RNAs: lessons from the Pyrococcus genomes. J Mol Biol 297: 895-906.
50. Omer AD, Lowe TM, Russell AG, Ebhardt H, Eddy SR, et al. (2000) Homologs of small nucleolar RNAs in Archaea. Science 288: 517-522.
51. Davila Lopez M, Rosenblad MA, Samuelsson T (2008) Computational screen for spliceosomal RNA genes aids in defining the phylogenetic distribution of major and minor spliceosomal components. Nucleic Acids Res 36: 3001-3010.
52. Russell AG, Charette JM, Spencer DF, Gray MW (2006) An early evolutionary origin for the minor spliceosome. Nature 443: 863-866.
53. Collins L, Penny D (2005) Complex spliceosomal organization ancestral to extant eukaryotes. Molecular Biology and Evolution 22: 1053-1066.
54. Shabalina SA, Koonin EV (2008) Origins and evolution of eukaryotic RNA interference. Trends Ecol Evol 23: 578-587.
55. Wu D, Hugenholtz P, Mavromatis K, Pukall R, Dalin E, et al. (2009) A phylogeny-driven genomic encyclopaedia of Bacteria and Archaea. Nature 462: 1056-1060.
56. Anantharaman V, Koonin EV, Aravind L (2002) Comparative genomics and evolution of proteins involved in RNA metabolism. Nucleic Acids Res 30: 1427-1464.
57. Cox CJ, Foster PG, Hirt RP, Harris SR, Embley TM (2008) The archaebacterial origin of eukaryotes. Proceedings of the National Academy of Sciences of the United States of America 105: 20356-20361.
58. Gribaldo S, Poole AM, Daubin V, Forterre P, Brochier-Armanet C (2010) The origin of eukaryotes and their relationship with the Archaea: are we at a phylogenomic impasse? Nat Rev Micro 8: 743-752.
59. Cochrane G, Akhtar R, Bonfield J, Bower L, Demiralp F, et al. (2009) Petabyte-scale innovations at the European Nucleotide Archive. Nucleic Acids Res 37: D19-25.
60. Brochier-Armanet C, Boussau B, Gribaldo S, Forterre P (2008) Mesophilic Crenarchaeota: proposal for a third archaeal phylum, the Thaumarchaeota. Nature reviews microbiology 6: 245-252.
61. Adl SM, Simpson AG, Farmer MA, Andersen RA, Anderson OR, et al. (2005) The new higher level classification of eukaryotes with emphasis on the taxonomy of protists. J Eukaryot Microbiol 52: 399-451.

62. Frickey T, Lupas A (2004) CLANS: a Java application for visualizing protein families based on pairwise similarity. Bioinformatics 20: 3702-3704.
63. Enright AJ, Van Dongen S, Ouzounis CA (2002) An efficient algorithm for large-scale detection of protein families. Nucleic Acids Res 30: 1575-1584.
64. Liu Y, Schmidt B, Maskell DL (2010) MSAProbs: multiple sequence alignment based on pair hidden Markov models and partition function posterior probabilities. Bioinformatics 26: 1958-1964.
65. Castresana J (2000) Selection of conserved blocks from multiple alignments for their use in phylogenetic analysis. Mol Biol Evol 17: 540-552.
66. Abascal F, Zardoya R, Posada D (2005) ProtTest: selection of best-fit models of protein evolution. Bioinformatics 21: 2104-2105.
67. Guindon S, Gascuel O (2003) A simple, fast, and accurate algorithm to estimate large phylogenies by maximum likelihood. Syst Biol 52: 696-704.
68. Stamatakis A (2006) RAxML-VI-HPC: maximum likelihood-based phylogenetic analyses with thousands of taxa and mixed models. Bioinformatics 22: 2688-2690.
69. Gascuel O (1997) BIONJ: an improved version of the NJ algorithm based on a simple model of sequence data. Mol Biol Evol 14: 685-695.
70. Huson DH, Richter DC, Rausch C, Dezulian T, Franz M, et al. (2007) Dendroscope: An interactive viewer for large phylogenetic trees. BMC Bioinformatics 8: 460.
71. Hartmann E, Hartmann RK (2003) The enigma of ribonuclease P evolution. Trends Genet 19: 561-569.
72. Fournier GP, Andam CP, Alm EJ, Gogarten JP (2011) Molecular Evolution of Aminoacyl tRNA Synthetase Proteins in the Early History of Life. Orig Life Evol Biosph.
73. Jacq B (1981) Sequence homologies between eukaryotic 5.8S rRNA and the 5' end of prokaryotic 23S rRNa: evidences for a common evolutionary origin. Nucleic Acids Res 9: 2913-2932.
74. Lafontaine DL, Tollervey D (2001) The function and synthesis of ribosomes. Nat Rev Mol Cell Biol 2: 514-520.

**Figure Legends**

**Fig. 1. Venn diagram of RNA family distribution.** Taxonomic information attached to EMBL-derived Rfam annotations reveals that the majority (99%) of RNA families are domain-specific, with only seven RNA families universally conserved (across the three domains of life plus viruses; **Table S1**). Numbers within dashed circles indicate viral RNA families.

**Fig. 2. RNA-based processes traceable to the Last Universal Common Ancestor.** Universal Rfam families that show evidence of vertical inheritance (**Table S1**) are all associated with the processes of translation (rRNAs, tRNAs, RNase P) and protein export (SRP RNA). A previous study examining the antiquity of protein coding genes [21] identified only 37 universally distributed proteins which show evidence of vertical inheritance. The majority of these vertically inherited proteins are associated with translation and protein export; numbers of such proteins associated with each of the depicted processes is given in grey (original data are from Harris [21]). The proteins associated with RNase P are not universally conserved, with archaeal and eukaryotic RNase P proteins being unrelated to their bacterial counterparts [71]. While tRNA synthetases are universal, they have undergone ancient horizontal gene transfer events [72], which complicates establishing the timing of their origin.

**Fig. 3. Reconstruction of broadly distributed RNA repertoires for each domain, plus interdomain RNA families.** Colored bars at far right indicate normalized taxonomic abundance of each Rfam for major taxonomic groupings within each domain.

Horizontal traces (see text, **Table S1**) for interdomain families, are depicted as follows: general transfer patterns are given by dashed arrows; proposed HGT patterns for individual families are depicted by number (inset). For Rfam families present in more than one domain (far left and inset), bars indicate normalized taxonomic abundance by domain (color scheme at bottom left). Asterisks indicate additional broadly-distributed bacterial candidates identified using GEBA tree topology [55] (see text). Note that the Rfam rRNA families in Rfam 10.0 are based on conserved subsequences, and are not as comprehensive as other resources (see main text) and are included here for consistency. The universally-distributed rRNAs are the small subunit (16/18S) rRNA, large subunit (23/28S) rRNA and 5S rRNA (see **Table S1**). The 5.8S rRNA of eukaryotes is known to be homologous to the 5' end of bacterial and archaeal 23S rRNA [73,74], so its inclusion as a eukaryote-specific family in Rfam is in this respect artefactual.

**Fig. 4. Rfam-based functional classification of RNA families.** The tree depicts classification of the higher level data structures within Rfam, and is not a phylogeny. Numbers of sequences and families in Rfam 10 that fall into each functional classification are shown as bar charts. Domain-level taxonomic distribution for each functional category is shown by black (present) and white (absent) boxes, right. The grey box indicates that H/ACA family RNAs are known from archaea [46,47], but are not in Rfam 10.

**Figure S1. Unrooted PhyML phylogeny of TPP-regulated gene product THIC.**

(A) Tree in landscape format so labels are legible. The phylogeny shows good support for a close affinity between Plant and green algal (green) and a clan of proteobacterial homologs (red), to the exclusion of archaeal sequences (dark blue), consistent with possible HGT from bacteria to eukaryotes. Monophyletic groups are not recovered for either archaea or bacteria, suggestive of horizontal transmission events. All tips are labeled with the following information: MCL_cluster|Domain|gi_number|species_name. Bootstrap values are out of 108 (Materials and Methods). (B) Same tree in unrooted form; coloring is identical to key in (A).

**Figure S2. Analysis of taxonomic distribution of Rfam entries within the EMBL nucleotide database.**

Data for each of the three domains (A) Eukarya (B) Archaea (C) Bacteria are binned by indicated major taxonomic groupings (see Materials and Methods). The x-axis corresponds to individual Rfam entries. The majority of families are restricted to well-studied groups, revealing a strong bias in the underlying data, as previously seen for snoRNA families [48] and more generally for genome projects [55].

**Figure S3. Discovery curves for Rfam.**

These curves plot the oldest reliable electronic date (EMBL entry or publication) associated with a particular Rfam family. Domain distribution (1-domain, 2-domain or 3-domain) is based on current distributions. To generate discovery curves for all RNA families in Rfam 10.0 (which includes families built before January 2010), we extracted the oldest dates from the literature references contained in the corresponding Stockholm file and from the EMBL accessions – the oldest date of the two is plotted.

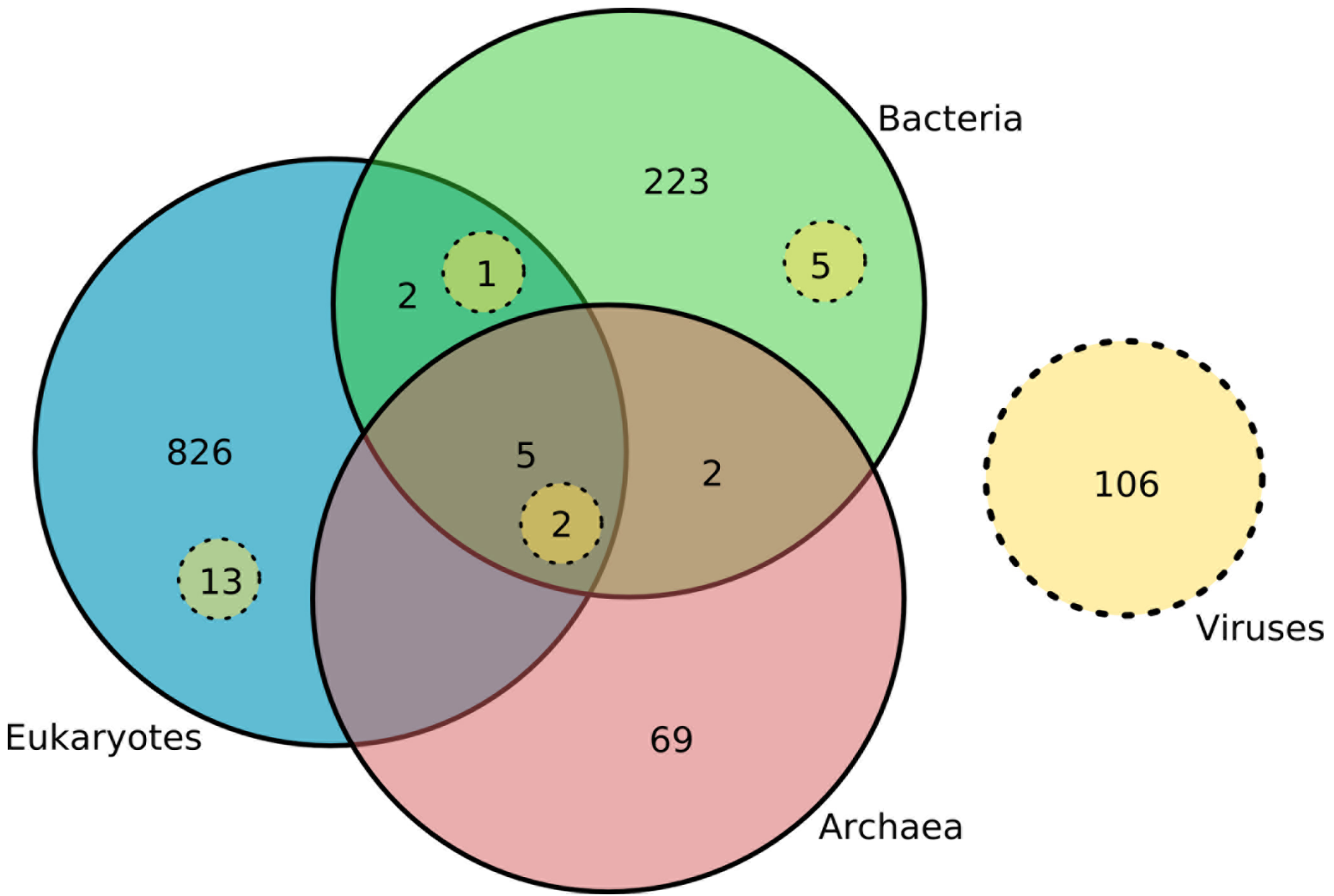

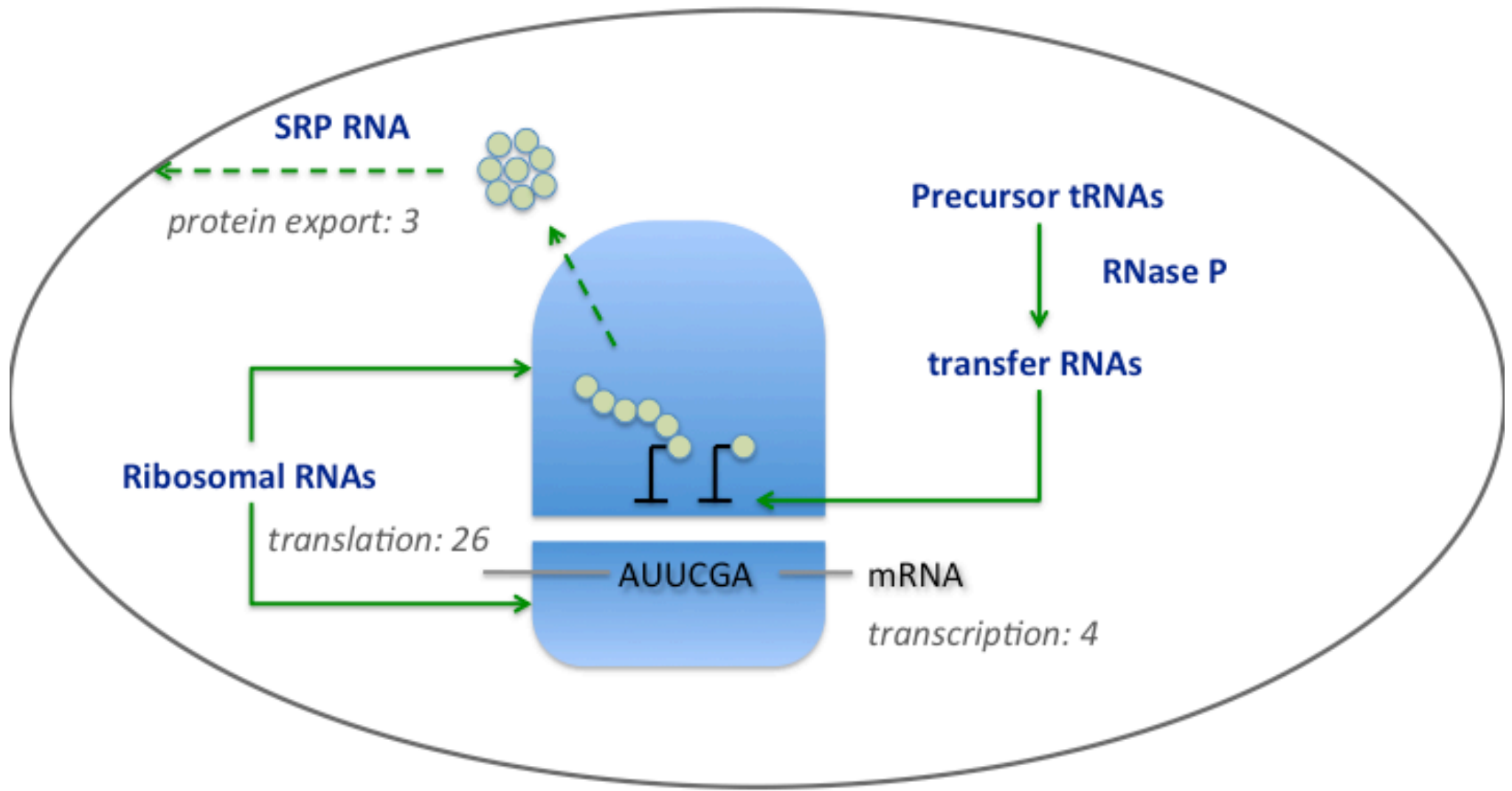

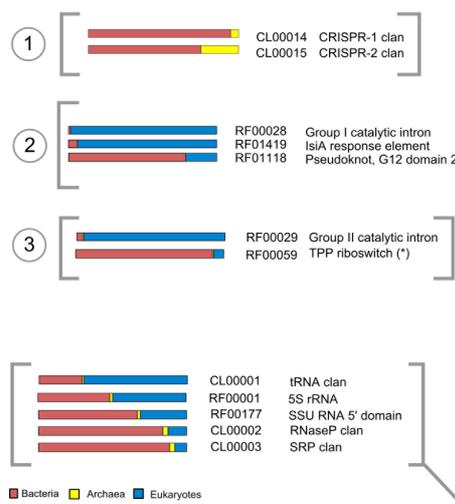
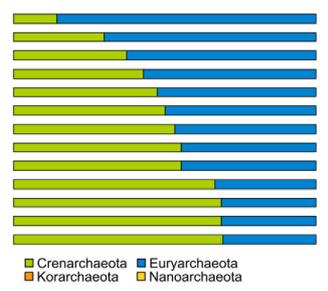
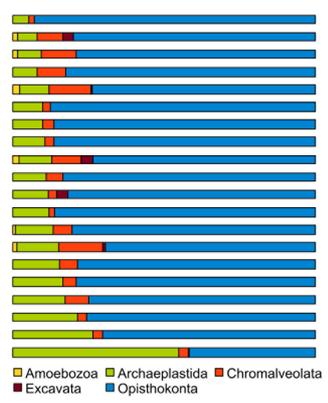
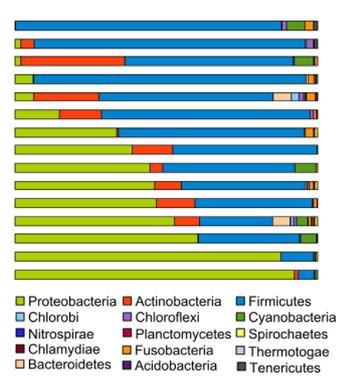

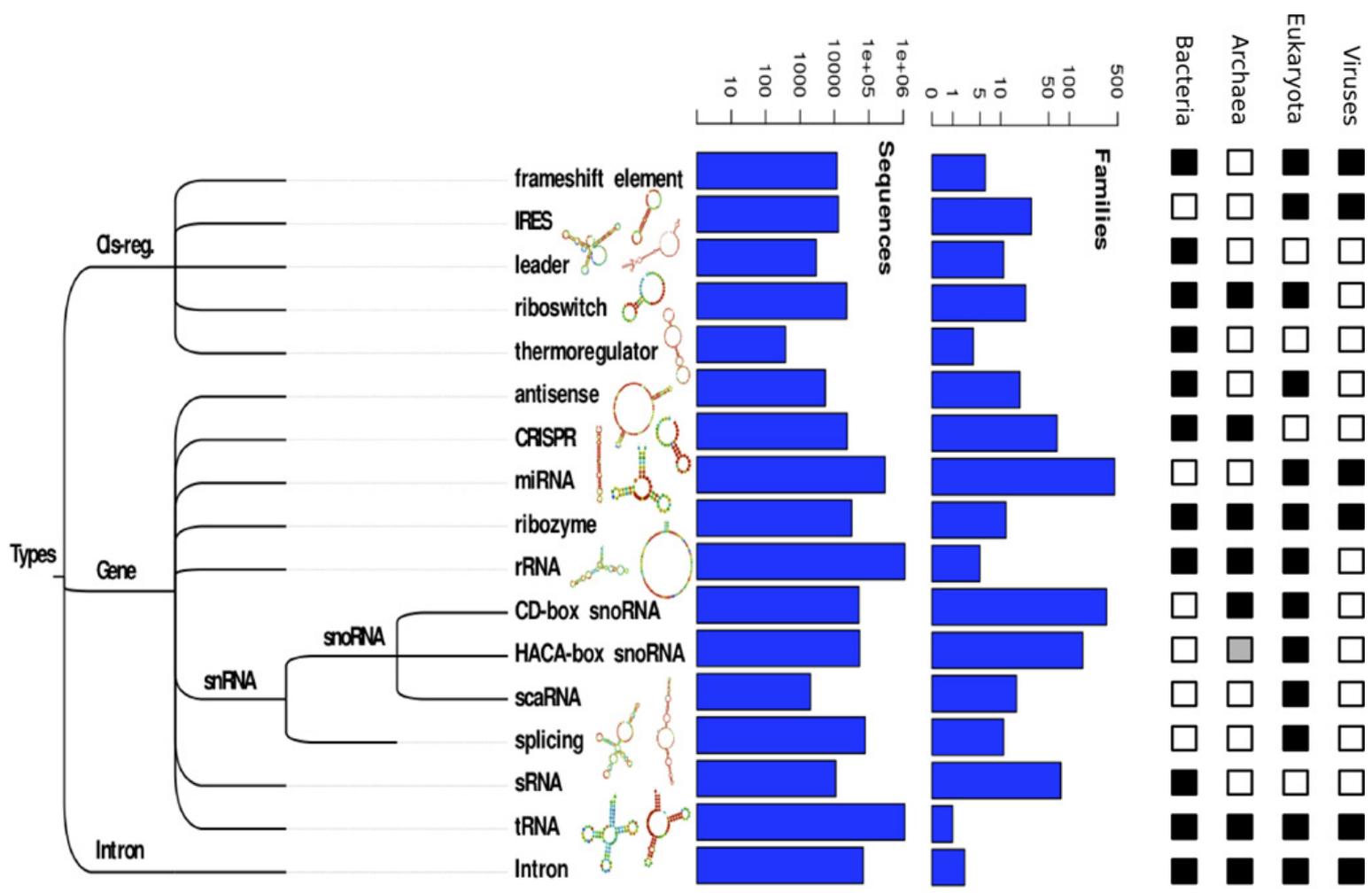

Supporting Information for

# Comparative analysis of RNA families reveals distinct repertoires for each domain of life

Marc P. Hoeppner[a], Paul P. Gardner[b] & Anthony M. Poole[b]*

[a]Science for Life Laboratory, Department of Medical Biochemistry and Microbiology, Uppsala University, SE-751 23 Uppsala, Sweden

[b]School of Biological Sciences, University of Canterbury, Private Bag 4800, Christchurch 8140, New Zealand

*To whom correspondence should be addressed.

Email addresses:

anthony.poole@canterbury.ac.nz

paul.gardner@canterbury.ac.nz

mphoeppner@gmail.com

**Supporting Information**

*Justification for the use of Rfam data for comparative analyses.*

There are several issues that must be considered for any analysis that requires homology inference. As a large scale comparative analysis of RNA families has not, to our knowledge, been performed before, it is important to begin such an analysis with a discussion of the merits — and possible limitations — of analyzing such a dataset across deep evolutionary history.

For both proteins and RNA, structure is often better conserved than sequence, such that homology may not be detectable from sequence data alone [1,2]. In Rfam, families are based on covariance models (CMs), which contain both primary sequence and secondary structure information [3,4], thus enabling detection of homology well below the twilight zone of sequence similarity for nucleic acids [2]. Evolutionary relationships for RNA genes from distant taxa have been reported (**Table S1**), and such distant similarities can be detected using Rfam [3] (**Hoeppner & Poole, submitted**), suggesting detection of homology is possible with a range of methods, even for distantly related RNAs.

Major classification schemes for protein families are based on measures of sequence similarity [5,6], which may fracture protein families classified from structure, where significant sequence similarity is undetectable (e.g. ribonucleotide reductases — [7]). If homology is routinely missed for deeply conserved RNAs, biologically-characterisable families should be artificially fractured. Examination of Rfam revealed few obvious cases, with those that we could identify being resolved at the clan level [3] (**Hoeppner & Poole, submitted**). In Rfam 10.0, 20% of families are further grouped into clans, and clan generation is achieved via implementation of a modified version of PRC [8], optimized for RNA profile:profile comparisions [3]. This permits detection of very distant relationships, based on both sequence and secondary structure similarity across multiple Rfam families. While no homology detection method can be claimed to be exhaustive, we believe Rfam is, in methodological terms, comparable to best practice in delineation of protein families by profile:profile comparisons [6,9,10]. For both types of data (RNA and protein), homology inference (i.e. defining families as a collection of sequences with a shared common ancestor) is made based on a measure of similarity.

There are known issues with existing sequence databases and datasets, where distinguishing between different forms of homology may be non-trivial, and is often not explicitly established [11,12]. As with protein families, Rfam families may therefore contain a mixture of orthologs (genes related by common descent following speciation), paralogs (genes related by common descent following gene duplication) and xenologs (genes related by common descent following horizontal gene transfer). For xenologs, this will not obviously affect the number of Rfam families, only their distribution. As discussed below, since this study focuses only on interdomain comparisons, we explicity examine xenologs at the level of interdomain comparisons, but not within domains. Within-domain xenology may impact the number of broadly-distributed RNAs. More generally, at the level of resolution used here (comparison across the three domains of life), failure to distinguish between orthologs and paralogs may at most alter the number of families attributed to each domain. Cases where paralogs are counted as separate families would artificially increase the number of within-domain families, and cases where functionally divergent paralogs are grouped within the same family would reduce this number. Mitigating against this, Rfam families and clans are based on a combination of sequence and structural similarity, plus common functionality [3], and inspection of clans indicates these represent orthologous groups rather than groupings of larger families with multiple paralogous constituents (personal observations). We think it is reasonable to conclude that the RNAs that make up individual Rfam families and clans can be considered to be homologous, and duly note that the caveats described here regarding orthology and paralogy apply equally to large protein-based datasets [12]. We can identify no sources of error that are demonstrably associated only with RNA data. For these reasons, we conclude that Rfam data is amenable to global comparative analyses.

*Rates of interdomain RNA family discovery.*
Given the currently rapid rate of discovery of novel RNAs, the Rfam database may not carry an up-to-the-minute picture of all known RNA families. The analysis we present is therefore necessarily a snapshot of current knowledge at the time of the Rfam release on which it is based, and will no doubt evolve as new RNA families are discovered. For the current study it is important to establish the rate of discovery of interdomain RNAs relative to intradomain RNAs. We therefore plotted discovery curves for all of Rfam (**Figure S3**). As is clear from **Figure S3**, the discovery rate of interdomain RNAs flattened off some time ago, whereas even in a conservative database like Rfam, domain-specific RNAs are still being added at a significant rate. There is no indication that interdomain RNAs lag far behind in terms of discovery. As is clear from **Figure S3** and **Table S4**, some newer cases even show a shortening of discovery times, perhaps because it is easier to screen for these in the post-genomic environment (e.g. [13]). We suspect that, as new data are published (as discussed in e.g. [14]), single-domain RNA families will continue to massively outpace discovery of new interdomain RNAs.

*Universally-distributed RNA families.*

Two families/clans show a universal distribution (present in all three domains plus viruses). For group II self-splicing introns, it is well established that these RNA elements are horizontally transmitted, with good evidence for recent transfer events from bacteria to archaea [15,16], and to eukaryotes via organelles [17]. By contrast, tRNAs, which are also universal, have been proposed to show a vertical evolutionary trace [18], and their involvement in viral replication has been argued to indicate an early evolutionary origin [19]. While individual tRNAs may have polyphyletic origins [20,21], placing presence of this family of RNAs in the ancestor of all three domains (**Figure 2**) is not controversial.

*RNA families present in all three domains.*

Five RNA families/clans are present in all three domains, and four of the five have been previously argued to show a vertical trace (**Table S1**). Rfam does not include full models for the large and small subunit ribosomal RNA, though RF00177 covers the 5' domain of the SSU rRNA. The only surprise member of this list is the TPP riboswitch. A difficulty with directly examining the evolutionary history of specific RNA elements in detail is that elements tend to be short, precluding reliable phylogenetic analysis in many cases. To abrogate this problem, we generated protein sequence phylogenies (Materials and Methods) derived from the most broadly distributed TPP-regulated gene, ThiC. With in excess of 4500 THIC sequences in genbank, we used MCL [22] to generate a broad overview of the data and selected representative sequences from each of the major MCL clusters for phylogenetic analysis (see Materials and Methods). Next, we performed phylogenetic analyses on a subset of this data, with sequence selection guided by the network of clusters (Materials and Methods). As is clear from **Figure S1**, we do not recover the monophyly of the three domains, with proteobacteria and archaea both split into distinct groups, which cannot be attributed to phylogenetic artefact. It therefore seems likely that the non-monophyly of both archaeal and bacterial ThiC sequences is best attributed to horizontal transmission events. Eukaryotes, in contrast, do form a single clan (sensu [23]). While the tree in **Figure S1** is unrooted, vertical descent of the eukaryote sequences from the Last Universal Common Ancestor is difficult to reconcile with the non-monophyly among the other two domains. This would require the position of the root to be between eukaryotes and bacteria/archaea. Given that the eukaryote sequences group with proteobacterial sequences, are relatively restricted in distribution, and surrounded by neighbouring bacterial clans, it seems more plausible that eukaryote ThiC sequences have entered this domain via horizontal gene transfer from a bacterial source during the evolution of the Archaeplastida.

*Interdomain RNA families.*

After vetting for false annotations, we recovered five additional families/clans with members present in more than one domain (**Table S1**). Evidence of horizontal transmission can be established for all five cases. **Table S2** shows that the distribution of CRISPR crRNA Rfam families is largely domain-specific, suggesting ongoing interdomain transfer is minimal. 53 of 65 families we analyzed are present in only a single domain. One family (RF01353) contains both archaeal and bacterial sequences but closer inspection reveals it to be archaeal-specific; this family carries only a single bacterial annotation, in Cyanothece sp. CCY0110, based on a single non-repeated, non-Cas associated region, making this almost certainly a false annotation. As noted in the main text, only two families show interdomain distribution, indicating at most limited interdomain transfer of crRNAs.

As per group II introns above, the horizontal transmission of group I self-splicing introns is well-documented (**Table S1**; [24]). The LSU rRNA pseudoknot is present in 23S rRNA from bacteria and eukaryotic organellar 23S rRNA; the latter entered eukaryotes via bacterial endosymbioses, as judged by representative 23S rRNA phylogenies, and congruence with 16S rRNA phylogenies [25]. All eukaryotic group II intron and 23S rRNA sequences annotated in the EMBL database were examined to establish their genomic location; in all cases, we find these are in organellar (chloroplast and mitochondrial) genomes. Finally, the IsrR iron stress repressed RNA is associated with photosystem I in the cyanobacterium *Synechocystis* sp. PCC 6803 [26], and annotated eukaryote sequences in the EMBL database are all chloroplast-encoded, strongly linking this element to the endosymbiotic origin of the chloroplast.

**Table S1.** Conservation of Rfam families and clans across domains & viruses

| Distribution | RNA | Rfam ID | Evolutionary trace |
|---|---|---|---|
| **Universal** | tRNA | CL00001 | Vertical [18,19] |
| | Group II intron | RF00029 | Horizontal: Bacteria to Archaea [15,16]; Bacteria to Eukaryotes via organelles [17]; examination of taxonomic distribution of eukaryotic group II introns annotated in EMBL (this study; see also http://www.rna.ccbb.utexas.edu/SAE/2C/) confirms all are encoded in chloroplast and mitochondrial genomes. |
| **3 domains** | Large and small subunit RNA | N/A | Vertical [25,27] |
| | SSU RNA, 5' domain | RF00177 | Vertical [27] |
| | 5S rRNA | RF00001 | Vertical [28] |
| | TPP riboswitch | RF00059 | Horizontal: this study (Fig. S2). |
| | RNase P RNA | CL00002 | Vertical [29,30] |
| | SRP RNA | CL00003 | Vertical [31,32] |
| **Prokaryotes** | crRNA: CRISPR-1 | CL00014 | Horizontal [33,34] |
| | crRNA: CRISPR-2 | CL00015 | Horizontal (as above) |
| **Viruses, Bacteria & Eukarya** | Group I intron | RF00028 | Horizontal: Bacteria to Eukaryotes via organelles [24]; examination of taxonomic distribution of eukaryotic group I introns annotated in EMBL (this study) confirms all are encoded in chloroplast and mitochondrial genomes. Note however that group I intron insertion into nuclear rRNA genes has also been described [24,35-37]; http://www.rna.ccbb.utexas.edu/SAE/2C/) |
| **Bacteria & Eukarya** | 23S rRNA Domain G (G12) pseudoknot | RF01118 | Vertical & horizontal: bacteria to eukaryotes [25]; examination of taxonomic distribution of eukaryotic 23S rRNAs annotated in EMBL (this study) confirms all are encoded in chloroplast and mitochondrial genomes. |
| | IsrR: Iron stress repressed RNA | RF01419 | Horizontal: bacteria to eukaryotes, photosystem I-associated in cyanobacteria [26]; distribution of eukaryotic IsrR RNAs annotated in EMBL (this study) confirms all are encoded in chloroplast genomes. |

**Table S2.** Distribution of CRISPR crRNA annotations in Rfam

| CLAN | RFAM ID | Bacteria Total species | Total RNAs | Archaea Total species | Total RNAs |
|---|---|---|---|---|---|
| CL00014 | RF01315 | 54 | 5037 | 4 | 123 |
| | RF01317 | 157 | 3367 | 0 | 0 |
| | RF01327 | 12 | 1019 | 0 | 0 |
| | RF01328 | 0 | 0 | 1 | 53 |
| | RF01338 | 0 | 0 | 7 | 528 |
| | RF01352 | 6 | 214 | 0 | 0 |
| | RF01379 | 1 | 128 | 0 | 0 |
| CL00015 | RF01318 | 12 | 608 | 0 | 0 |
| | RF01320 | 6 | 348 | 6 | 293 |
| | RF01376 | 1 | 23 | 0 | 0 |
| | RF01377 | 0 | 0 | 1 | 4 |
| Singleton families | RF01316 | 25 | 731 | 0 | 0 |
| | RF01319 | 0 | 0 | 7 | 615 |
| | RF01321 | 4 | 78 | 0 | 0 |
| | RF01322 | 8 | 197 | 0 | 0 |
| | RF01323 | 6 | 452 | 0 | 0 |
| | RF01324 | 0 | 0 | 1 | 38 |
| | RF01325 | 6 | 213 | 0 | 0 |
| | RF01326 | 0 | 0 | 2 | 139 |
| | RF01329 | 2 | 15 | 0 | 0 |
| | RF01330 | 2 | 51 | 0 | 0 |
| | RF01331 | 3 | 114 | 0 | 0 |
| | RF01332 | 5 | 1288 | 0 | 0 |
| | RF01333 | 2 | 73 | 0 | 0 |
| | RF01334 | 2 | 548 | 0 | 0 |
| | RF01335 | 53 | 518 | 0 | 0 |
| | RF01336 | 2 | 177 | 0 | 0 |
| | RF01337 | 0 | 0 | 8 | 289 |
| | RF01339 | 0 | 0 | 1 | 125 |
| | RF01340 | 1 | 24 | 0 | 0 |
| | RF01341 | 1 | 19 | 0 | 0 |
| | RF01342 | 1 | 30 | 0 | 0 |
| | RF01343 | 5 | 62 | 0 | 0 |
| | RF01344 | 25 | 74 | 0 | 0 |
| | RF01345 | 2 | 32 | 0 | 0 |
| | RF01346 | 4 | 158 | 0 | 0 |
| | RF01347 | 3 | 34 | 0 | 0 |
| | RF01348 | 5 | 120 | 0 | 0 |
| | RF01349 | 3 | 82 | 0 | 0 |
| | RF01350 | 0 | 0 | 2 | 118 |
| | RF01351 | 0 | 0 | 2 | 9 |
| | RF01353 | 1 | 1 | 2 | 69 |
| | RF01354 | 0 | 0 | 4 | 327 |
| | RF01355 | 0 | 0 | 4 | 203 |
| | RF01356 | 6 | 498 | 0 | 0 |
| | RF01357 | 2 | 37 | 0 | 0 |
| | RF01358 | 0 | 0 | 2 | 89 |
| | RF01359 | 1 | 6 | 0 | 0 |
| | RF01360 | 0 | 0 | 1 | 43 |
| | RF01361 | 2 | 11 | 0 | 0 |
| | RF01362 | 9 | 182 | 0 | 0 |
| | RF01363 | 2 | 48 | 0 | 0 |
| | RF01364 | 1 | 23 | 0 | 0 |
| | RF01365 | 1 | 50 | 0 | 0 |
| | RF01366 | 1 | 14 | 0 | 0 |
| | RF01367 | 1 | 15 | 0 | 0 |
| | RF01368 | 1 | 34 | 0 | 0 |
| | RF01369 | 0 | 0 | 2 | 92 |
| | RF01370 | 5 | 103 | 0 | 0 |
| | RF01371 | 2 | 139 | 0 | 0 |
| | RF01372 | 4 | 18 | 0 | 0 |
| | RF01373 | 0 | 0 | 3 | 171 |
| | RF01374 | 16 | 83 | 0 | 0 |
| | RF01375 | 0 | 0 | 1 | 29 |
| | RF01378 | 0 | 0 | 2 | 101 |

**Table S3.** Distribution of spliceosomal RNAs across eukaryotes*

|       | Amoebozoa | | Opisthonkonts | | Archaeplastida | | Chromalveolata | | Excavates | | | Rhizaria | |
|-------|------|----|------|----|------|----|------|----|------|----|------|------|----|
|       | Rfam | DL | Rfam | DL | Rfam | DL | Rfam | DL | Rfam | DL | Chen | Rfam | DL |
| U1    | 7    | x  | 10663| x  | 415  | x  | 85   | x  | 2    | x  | x    | 0    |    |
| U2    | 22   | x  | 6682 | x  | 540  | x  | 183  | x  | 30   | x  | x    | 0    |    |
| U4    | 4    | x  | 5369 | x  | 143  | x  | 61   | x  | 0    | x  | x    | 0    |    |
| U5    | 9    | x  | 2996 | x  | 330  | x  | 87   | x  | 1    | x  | x    | 0    |    |
| U6    | 9    | x  | 47847| x  | 356  | x  | 205  | x  | 19   | x  | x    | 0    |    |
| U11   | 0    | x  | 393  | x  | 29   | x  | 4    | x  | 0    |    |      | 0    |    |
| U12   | 1    | x  | 295  | x  | 22   | x  | 4    | x  | 0    |    |      | 0    |    |
| U4atac| 0    |    | 419  | x  | 0    | x  | 0    |    | 0    |    |      | 0    |    |
| U6atac| 0    |    | 1560 | x  | 60   | x  | 4    | x  | 0    |    |      | 0    |    |

*Data are derived from this study (Rfam), Davila-Lopez et al. (DL)[38] and Chen et al. (Chen)[39]. Crosses (x) denote presence in one or more eukaryote species within the supergroup; Rfam counts are total number of annotations in EMBL, release 100.

**Table S4.** Records associated with RNA families in the Rfam database[a]

| Rfam | EMBL | PUB DATE | DESCRIPTION | Conservation[b] |
|---|---|---|---|---|
| RF00028 | 2004 | 1990 | Group I catalytic intron | E-B |
| RF01118 | 2008 | 1987 | Pseudoknot of the domain G(G12) of 23S ribosomal RNA | E-B |
| RF01419 | 1989 | 2006 | Antisense RNA which regulates isiA expression | E-B |
| RF01317 | 2006 | N/A | CRISPR RNA direct repeat element | A-B |
| RF01338 | 2007 | N/A | CRISPR RNA direct repeat element | A-B |
| RF00001 | 1992 | 2000 | 5S ribosomal RNA | LUCA |
| RF00002 | 1993 | 1997 | 5.8S ribosomal RNA | LUCA |
| RF00005 | 1994 | 1993 | tRNA | LUCA |
| RF00009 | 1996 | 1998 | Nuclear RNase P | LUCA |
| RF00010 | 1986 | 1998 | Bacterial RNase P class A | LUCA |
| RF00011 | 1996 | 1998 | Bacterial RNase P class B | LUCA |
| RF00017 | 2005 | 2000 | Eukaryotic type signal recognition particle RNA | LUCA |
| RF00023 | 2006 | 1996 | transfer-messenger RNA | LUCA |
| RF00029 | 2002 | 2001 | Group II catalytic intron | LUCA |
| RF00030 | 2005 | 1993 | RNase MRP | LUCA |
| RF00059 | 2007 | 2001 | TPP riboswitch (THI element) | LUCA |
| RF00169 | 1995 | 2002 | Bacterial signal recognition particle RNA | LUCA |
| RF00177 | 1991 | N/A | Small subunit ribosomal RNA, 5' domain | LUCA |
| RF00373 | 1991 | 1998 | Archaeal RNase P | LUCA |

[a]Displayed are the oldest dates from the literature references (PUB DATE) contained in the corresponding Stockholm file and from the EMBL accessions. Some of these RNAs were discovered experimentally prior to the dates associated with the deposited sequences, meaning the age of many of the oldest RNAs is in fact underestimated. Consequently, the discovery dates summarized in Fig. S3 are estimates that can only be used in the context of broad discovery trends.

[b]Abbreviations: present in eukaryotes & bacteria (E-B); present in archaea & bacteria (A-B); present in archaea, bacteria & eukaryotic domains (LUCA).

**Figure S1. Unrooted PhyML phylogeny of TPP-regulated gene product THIC.**
(A) Tree in landscape format so labels are legible. The phylogeny shows good support for a close affinity between Plant and green algal (green) and a clan of proteobacterial homologs (red), to the exclusion of archaeal sequences (dark blue), consistent with possible HGT from bacteria to eukaryotes. Monophyletic groups are not recovered for either archaea or bacteria, suggestive of horizontal transmission events. All tips are labeled with the following information: MCL_cluster|Domain|gi_number|species_name. Bootstrap values are out of 108 (Materials and Methods). (B) Same tree in unrooted form; coloring is identical to key in (A).

**Figure S2. Analysis of taxonomic distribution of Rfam entries within the EMBL nucleotide database.**
Data for each of the three domains (A) Eukarya (B) Archaea (C) Bacteria are binned by indicated major taxonomic groupings (see Materials and Methods). The x-axis corresponds to individual Rfam entries. The majority of families are restricted to well-studied groups, revealing a strong bias in the underlying data, as previously seen for snoRNA families [40] and more generally for genome projects [41].

**Figure S3. Discovery curves for Rfam.**
These curves plot the oldest reliable electronic date (EMBL entry or publication) associated with a particular Rfam family. Domain distribution (1-domain, 2-domain or 3-domain) is based on current distributions. To generate discovery curves for all RNA families in Rfam 10.0 (which includes families built before January 2010), we extracted the oldest dates from the literature references contained in the corresponding Stockholm file and from the EMBL accessions – the oldest date of the two is plotted.

# References

1. Illergard K, Ardell DH, Elofsson A (2009) Structure is three to ten times more conserved than sequence--a study of structural response in protein cores. Proteins 77: 499-508.
2. Freyhult EK, Bollback JP, Gardner PP (2007) Exploring genomic dark matter: a critical assessment of the performance of homology search methods on noncoding RNA. Genome Res 17: 117-125.
3. Gardner PP, Daub J, Tate J, Moore BL, Osuch IH, et al. (2011) Rfam: Wikipedia, clans and the "decimal" release. Nucleic Acids Res 39: D141-145.
4. Durbin R, Eddy SR, Krog A, Mitchison G (1998) Biological Sequence Analysis: Probabilistic Models of Proteins and Nucleic Acids: Cambridge University Press.
5. Tatusov RL, Natale DA, Garkavtsev IV, Tatusova TA, Shankavaram UT, et al. (2001) The COG database: new developments in phylogenetic classification of proteins from complete genomes. Nucleic Acids Res 29: 22-28.
6. Finn RD, Mistry J, Tate J, Coggill P, Heger A, et al. (2010) The Pfam protein families database. Nucleic Acids Res 38: D211-222.
7. Sintchak MD, Arjara G, Kellogg BA, Stubbe J, Drennan CL (2002) The crystal structure of class II ribonucleotide reductase reveals how an allosterically regulated monomer mimics a dimer. Nature structural biology 9: 293-300.
8. Madera M (2008) Profile Comparer: a program for scoring and aligning profile hidden Markov models. Bioinformatics 24: 2630-2631.
9. Cuff AL, Sillitoe I, Lewis T, Clegg AB, Rentzsch R, et al. (2011) Extending CATH: increasing coverage of the protein structure universe and linking structure with function. Nucleic Acids Res 39: D420-426.
10. Andreeva A, Howorth D, Chandonia JM, Brenner SE, Hubbard TJ, et al. (2008) Data growth and its impact on the SCOP database: new developments. Nucleic Acids Res 36: D419-425.
11. Fitch WM (2000) Homology a personal view on some of the problems. Trends Genet 16: 227-231.
12. Altenhoff AM, Dessimoz C (2009) Phylogenetic and functional assessment of orthologs inference projects and methods. PLoS Comput Biol 5: e1000262.
13. Weinberg Z, Wang JX, Bogue J, Yang J, Corbino K, et al. (2010) Comparative genomics reveals 104 candidate structured RNAs from bacteria, archaea, and their metagenomes. Genome Biol 11: R31.
14. Haas BJ, Zody MC (2010) Advancing RNA-Seq analysis. Nat Biotechnol 28: 421-423.
15. Rest JS, Mindell DP (2003) Retroids in archaea: phylogeny and lateral origins. Mol Biol Evol 20: 1134-1142.
16. Dai L, Zimmerly S (2003) ORF-less and reverse-transcriptase-encoding group II introns in archaebacteria, with a pattern of homing into related group II intron ORFs. RNA 9: 14-19.
17. Lambowitz AM, Zimmerly S (2004) Mobile group II introns. Annu Rev Genet 38: 1-35.
18. Sun FJ, Caetano-Anolles G (2008) Evolutionary patterns in the sequence and structure of transfer RNA: early origins of archaea and viruses. PLoS Comput Biol 4: e1000018.

19. Maizels N, Weiner AM (1999) The genomic tag hypothesis: what molecular fossils tell us about the evolution of tRNA. In: Gesteland RF, Cech TR, Atkins JF, editors. The RNA world, 2nd ed. 2nd ed ed. Cold Spring Harbor, NY: Cold Spring Harbor Laboratory Press. pp. 79–111.
20. Di Giulio M (1999) The non-monophyletic origin of the tRNA molecule. J Theor Biol 197: 403-414.
21. Rodin AS, Szathmary E, Rodin SN (2011) On origin of genetic code and tRNA before translation. Biology Direct: in press.
22. Enright AJ, Van Dongen S, Ouzounis CA (2002) An efficient algorithm for large-scale detection of protein families. Nucleic Acids Res 30: 1575-1584.
23. Wilkinson M, McInerney JO, Hirt RP, Foster PG, Embley TM (2007) Of clades and clans: terms for phylogenetic relationships in unrooted trees. Trends Ecol Evol 22: 114-115.
24. Haugen P, Simon DM, Bhattacharya D (2005) The natural history of group I introns. Trends Genet 21: 111-119.
25. Cedergren R, Gray MW, Abel Y, Sankoff D (1988) The evolutionary relationships among known life forms. J Mol Evol 28: 98-112.
26. Duhring U, Axmann IM, Hess WR, Wilde A (2006) An internal antisense RNA regulates expression of the photosynthesis gene isiA. Proc Natl Acad Sci USA 103: 7054-7058.
27. Woese CR, Kandler O, Wheelis ML (1990) Towards a natural system of organisms: proposal for the domains Archaea, Bacteria, and Eucarya. Proc Natl Acad Sci USA 87: 4576-4579.
28. Sun FJ, Caetano-Anolles G (2009) The evolutionary history of the structure of 5S ribosomal RNA. J Mol Evol 69: 430-443.
29. Sun FJ, Caetano-Anolles G (2010) The ancient history of the structure of ribonuclease P and the early origins of Archaea. BMC Bioinformatics 11: 153.
30. Collins LJ, Moulton V, Penny D (2000) Use of RNA secondary structure for studying the evolution of RNase P and RNase MRP. Journal of molecular evolution 51: 194-204.
31. Schmitz U, Behrens S, Freymann DM, Keenan RJ, Lukavsky P, et al. (1999) Structure of the phylogenetically most conserved domain of SRP RNA. RNA 5: 1419-1429.
32. Zwieb C, van Nues RW, Rosenblad MA, Brown JD, Samuelsson T (2005) A nomenclature for all signal recognition particle RNAs. RNA (New York, NY 11: 7-13.
33. Kunin V, Sorek R, Hugenholtz P (2007) Evolutionary conservation of sequence and secondary structures in CRISPR repeats. Genome Biol 8: R61.
34. Shah SA, Garrett RA (2011) CRISPR/Cas and Cmr modules, mobility and evolution of adaptive immune systems. Res Micro 162: 27-38.
35. Nikoh N, Fukatsu T (2001) Evolutionary dynamics of multiple group I introns in nuclear ribosomal RNA genes of endoparasitic fungi of the genus Cordyceps. Mol Biol Evol 18: 1631-1642.
36. Cannone JJ, Subramanian S, Schnare MN, Collett JR, D'Souza LM, et al. (2002) The comparative RNA web (CRW) site: an online database of comparative sequence and structure information for ribosomal, intron, and other RNAs. BMC Bioinformatics 3: 2.
37. Perotto S, Nepote-Fus P, Saletta L, Bandi C, Young JP (2000) A diverse population of introns in the nuclear ribosomal genes of ericoid mycorrhizal fungi includes elements with sequence similarity to endonuclease-coding genes. Mol Biol Evol 17: 44-59.
38. Davila Lopez M, Rosenblad MA, Samuelsson T (2008) Computational screen for spliceosomal RNA genes aids in defining the phylogenetic distribution of major and minor spliceosomal components. Nucleic Acids Res 36: 3001-3010.
39. Chen XS, White WT, Collins LJ, Penny D (2008) Computational identification of four spliceosomal snRNAs from the deep-branching eukaryote Giardia intestinalis. PLoS One 3: e3106.
40. Gardner PP, Bateman A, Poole AM (2010) SnoPatrol: how many snoRNA genes are there? J Biol 9: 4.
41. Wu D, Hugenholtz P, Mavromatis K, Pukall R, Dalin E, et al. (2009) A phylogeny-driven genomic encyclopaedia of Bacteria and Archaea. Nature 462: 1056-1060.

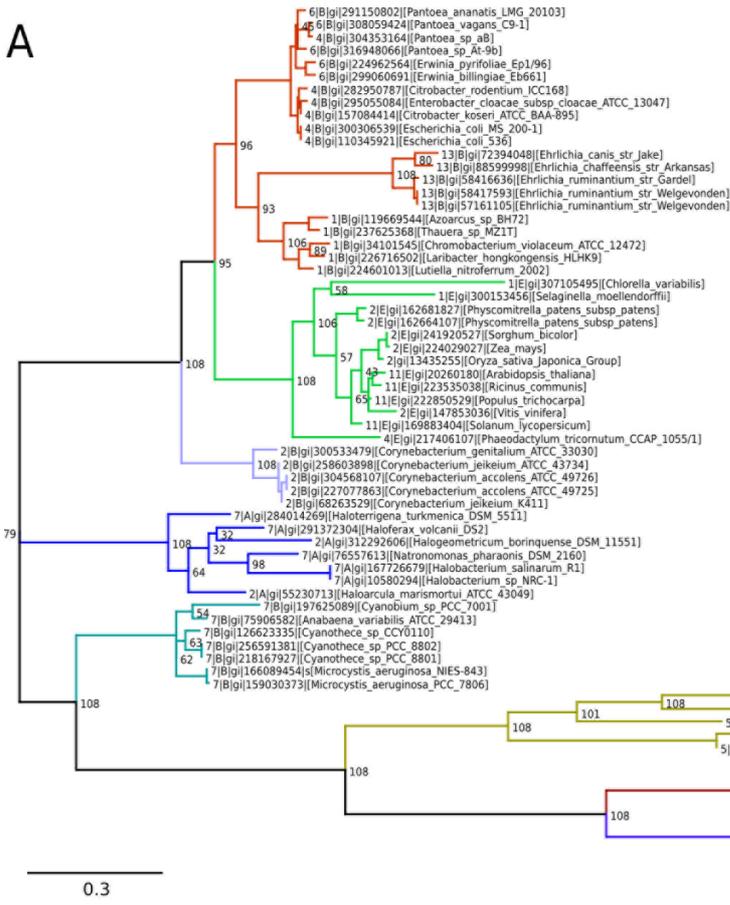
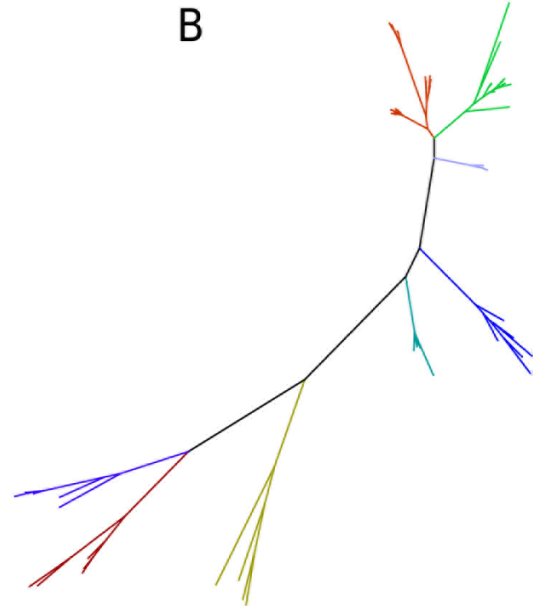

| Plants and green algae | Proteobacteria | Firmicutes | Actinobacteria | Cyanobacteria | Archaea |

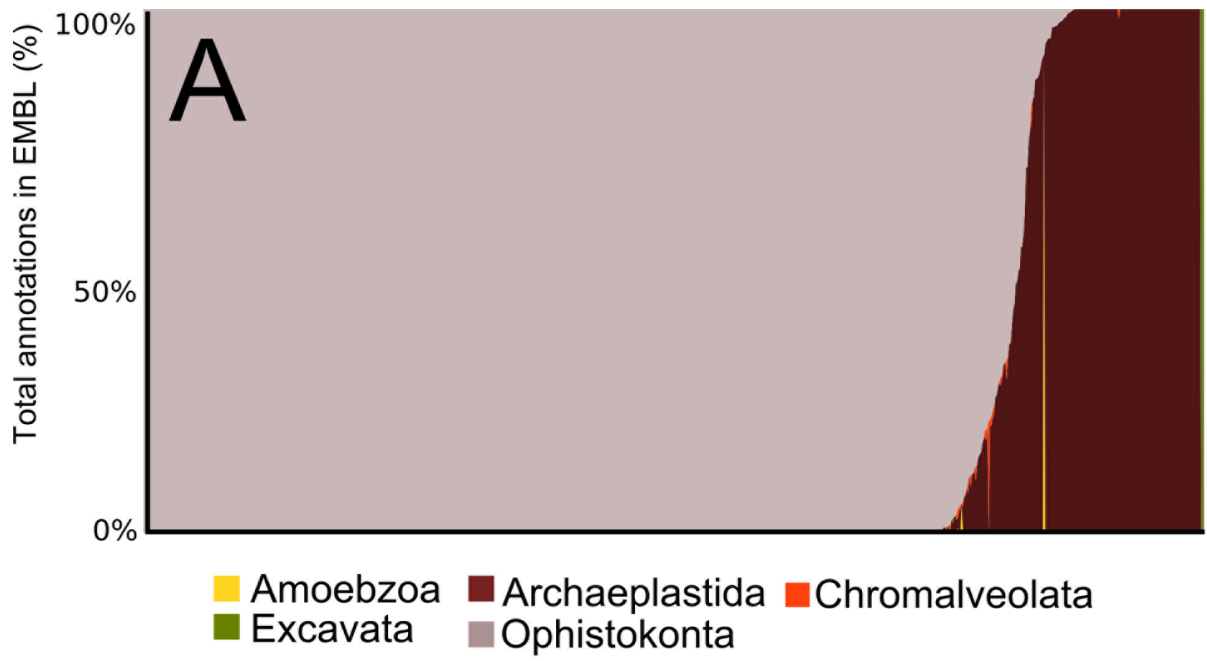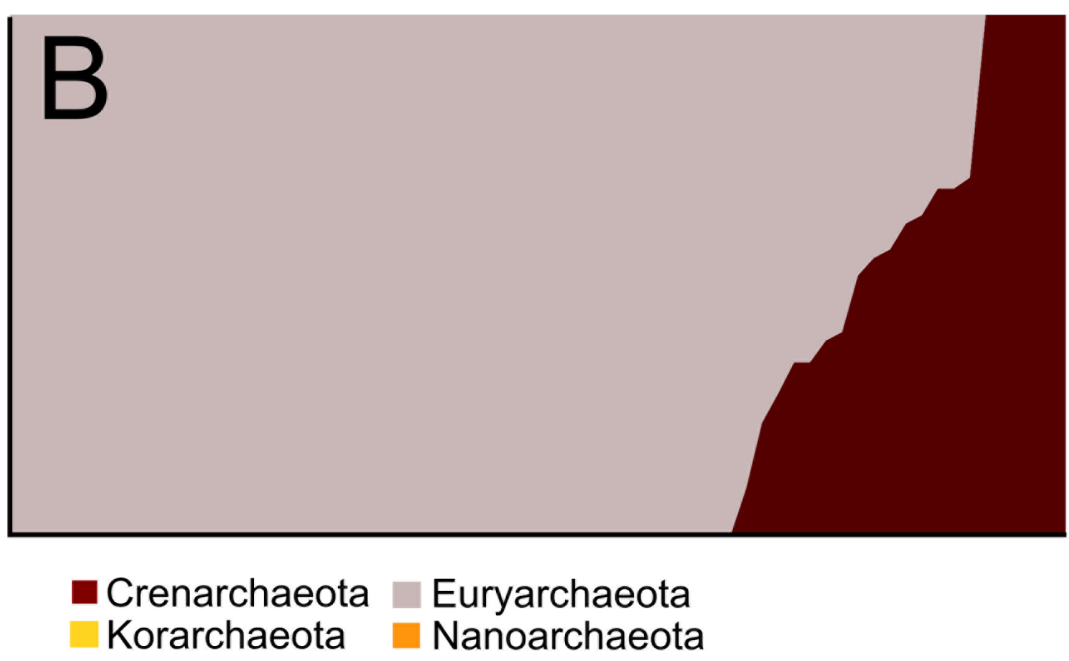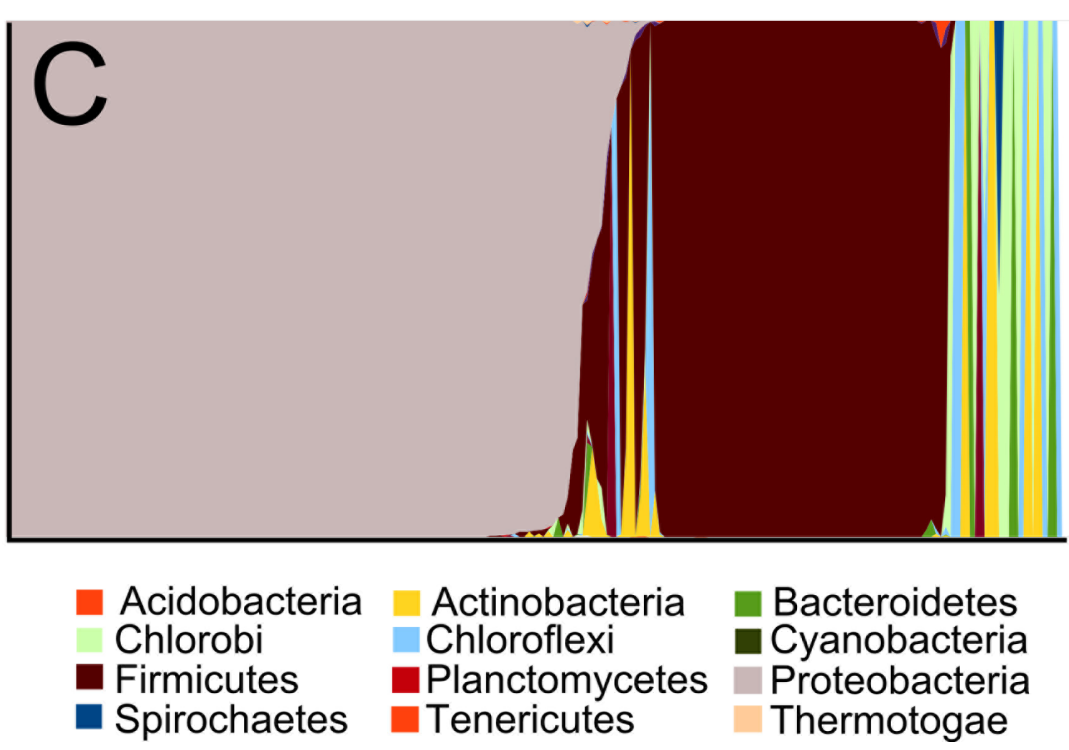

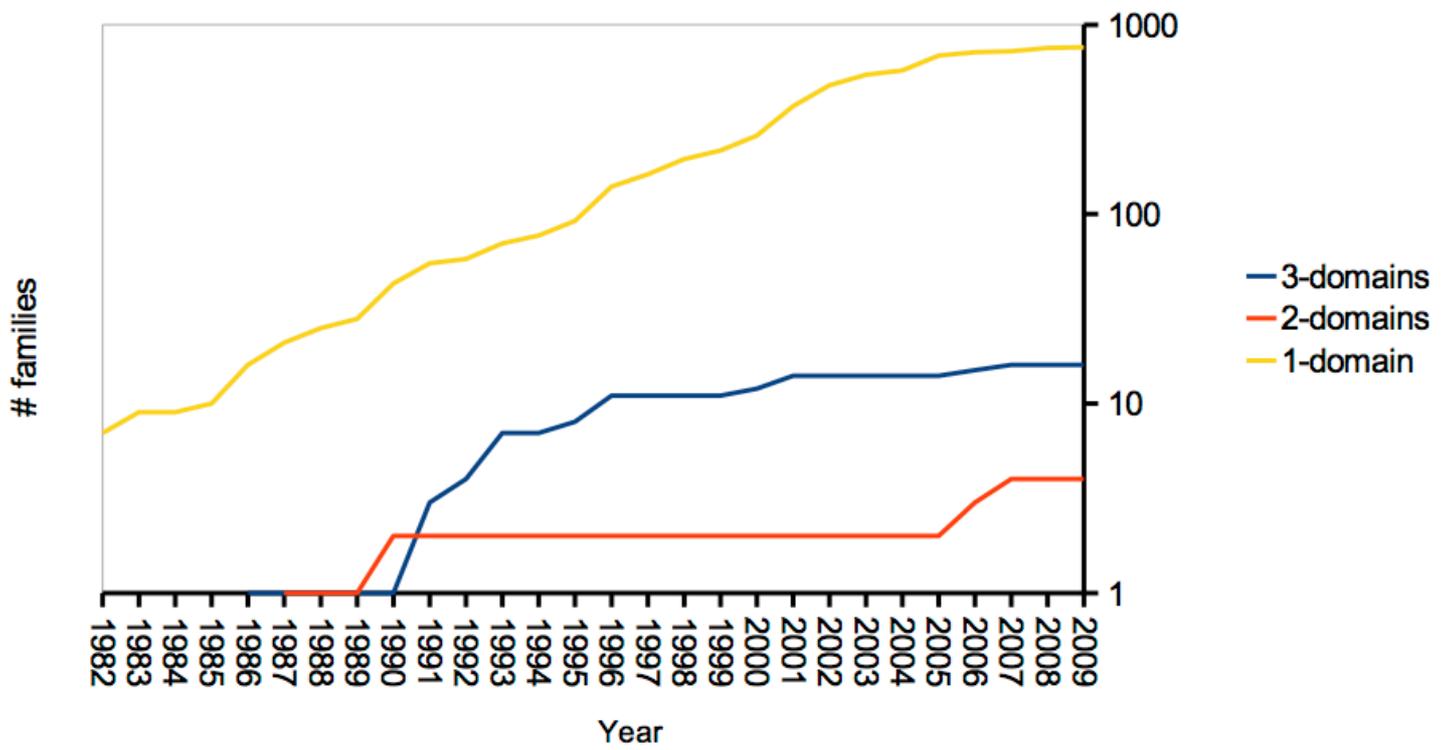